\documentclass[12pt]{iopart}

\usepackage{graphicx}
\usepackage[utf8]{inputenc}
\usepackage[english]{babel}
\usepackage[square,sort&compress,numbers]{natbib}
\newcommand{\be}{\begin{equation}}
\newcommand{\ee}{\end{equation}}
\newcommand{\om}{\Omega_M}
\newcommand{\ol}{\Omega_\Lambda}
\newcommand{\ok}{\Omega_k}
\newcommand{\gesim}{\,\raisebox{-0.4ex}{$\stackrel{>}{\scriptstyle\sim}$}\,}
\newcommand{\lesim}{\,\raisebox{-0.4ex}{$\stackrel{<}{\scriptstyle\sim}$}\,}

\begin{document}
\title[The universe as a particle]
{Cosmological histories from the Friedmann equation: The universe
as a particle}
\author{Edvard M{\"o}rtsell$^{1,2}$}
\address{$^1$ Oskar Klein Centre, Stockholm University, AlbaNova University Center, 106 91 Stockholm, Sweden}
\address{$^2$ Department of Physics, Stockholm University, AlbaNova University Center, 106 91 Stockholm, Sweden}
\ead{edvard@fysik.su.se}
\begin{abstract}
In this note, we discuss how possible expansion histories of the universe can be inferred in a simple way, 
for arbitrary energy contents. 
No new physical results are obtained, but the goal is rather to discuss an alternative way of writing the Friedmann 
equation in order to facilitate an intuitive understanding of the possible solutions; for students and researchers alike. 
As has been noted in passing by others, this specific form of the Friedmann equation allows us to view the universal expansion as a particle rolling along a frictionless track. Specific examples depicted include the current concordance cosmological model as well as a stable static universal model.
\end{abstract}
\submitto{\EJP}


\section{Introduction}\label{sec:intro}

In 1917, Albert Einstein \cite{Einstein:1917ce} and Willem de Sitter \cite{deSitter:1917zz} both suggested that our Universe could be described in terms of the relativistic field equations proposed by Einstein two years earlier \cite{Einstein:1915ca}. Being guided by the principle that inertia could be defined only in relation to other matter sources, Einstein's model had finite spatial extent and introduced the cosmological constant in order to achieve a static (background) distribution of matter. de Sitter on the other hand avoided the assumption of the universe being static since ``we only have a snapshot of the world, and we cannot and must not
conclude (\ldots) that everything will always remain as at that instant when the picture was
taken'' \cite{Realdi}. Instead, de Sitter's model was devoid of matter and any test particle initially at rest with respect to an observer would not remain so but rather require a positive radial velocity. At this point, the choice between Einstein's model (with matter but no motion) and de Sitter's model (with motion but no matter) was purely a matter of taste. 

In 1927, a non-static solution to the field equations, including matter sources, was proposed by Georges Lema\^{i}tre 
\cite{Lemaitre:1927zz}. Taken into account the possibility of the radius, or scale factor, of the universe to depend
on time, it was clear that Einstein's model was not stable since any small deviation from the equilibrium values of the energy densities would cause the universe to grow or contract. Being used to a time evolving scale factor as we are today, it may appear surprising that Einstein did not at first realize the instability of his proposed static solution. The reason for this of course that when not allowing for such a time dependence, the solution is perfectly well-behaved.
For the first time, Lema\^{i}tre also related the solution to the then available observations of recession velocities of distant sources, today interpreted as evidence for the expansion of the Universe.

 What is the reason then for us today referring to the corresponding equations as the Friedmann equations? Already in 1922, Alexander Friedmann presented solutions to the field equations, with a time dependent scale factor, including matter \cite{Friedmann:1924bb} (English translation). Even though Friedmann's work was both refuted and later unrefuted by Einstein himself, the solutions were note fully acknowledged until a few years after Lema\^{i}tre's rediscovery.

The history of the observational situation, and corresponding interpretations, regarding the universal dynamics is more unclear than the theoretical, as discussed in e.g. \cite{Way:2013ky} and references therein. Without entering this debate, we simply acknowledge the work of Vesto Slipher, Milton Humason, Henrietta Swan Leavitt, Knut Lundmark, George Lema\^{i}tre and Edwin Hubble, and note that in the thirties, it was observationally proved that the Universe was expanding and that this was a natural outcome of general relativity. 

In our current understanding of the Friedmann equations, we know that the expansion velocity today is not fixed by theory\footnote{Unless we have complete knowledge of both the energy densities and the curvature of the universe.} but that the evolution of this velocity, i.e. the acceleration, is set by the energy content of the universe. An empty universe will expand with constant velocity, whereas pressureless matter decelerates the expansion and a cosmological constant gives accelerated expansion. Einstein's static solution to the field equations is accomplished by having exact specific amounts of these counteracting energy components. The effect on the expansion velocity from an energy component is set by the relation between its density $\rho$ and pressure $p$, or the {\em equation of state} $\omega$, defined for a perfect fluid as
\be
p\equiv\omega c^2\rho.
\ee
The limiting case for an energy component to accelerate or decelerate the universal expansion is $p = -\rho c^2/3$, or $\omega=-1/3$ (see equation~\ref{eq:acc}). An energy component with this equation of state does not effect the expansion rate. However, it will still affect observations since it will have an effect on the spatial geometry, which in turn affects distance measures.

Given the total energy content of a universe, given as a sum of different perfect fluids such as radiation, pressureless matter, a cosmological constant etc, it is not obvious what kind of expansion histories are possible, e.g. if the model has a Big Bang, if it will expand forever, if it accelerates etc. Analytical solutions to the Friedmann equations are in principle only useful when one energy component dominates, e.g. in the early Universe. Although analytical solutions exist also in other cases, for most practical reasons they are not used today since numerical integration is generally favoured because it is fast, simple and can be used for arbitrary energy contents. Due to their often complicated structure, their use as a pedagogical tool is also limited. In this note, it is argued that a mechanical picture of the universal expansion involving kinetic and potential energies, instead can give a simple and full understanding of the qualitative structure of solutions. Since this only requires a very simple rewriting of the Friedmann equation from its most common form, this rewriting is not novel to this paper, see e.g. \cite{Szydlowski:2006ma,Padmanabhan:2006kz,Sonego:2011rb}. The purpose of this note is to clarify how the mechanical picture can be used to acquire an intuitive picture of the possible expansion histories of homogeneous and isotropic universes in general relativity.

\begin{figure}
\includegraphics[width=8cm]{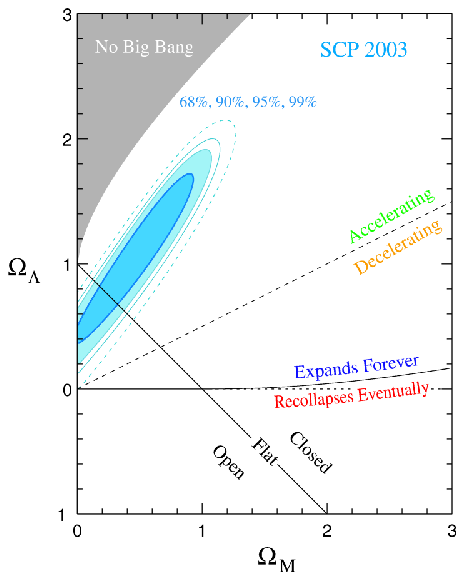}\includegraphics[width=8cm]{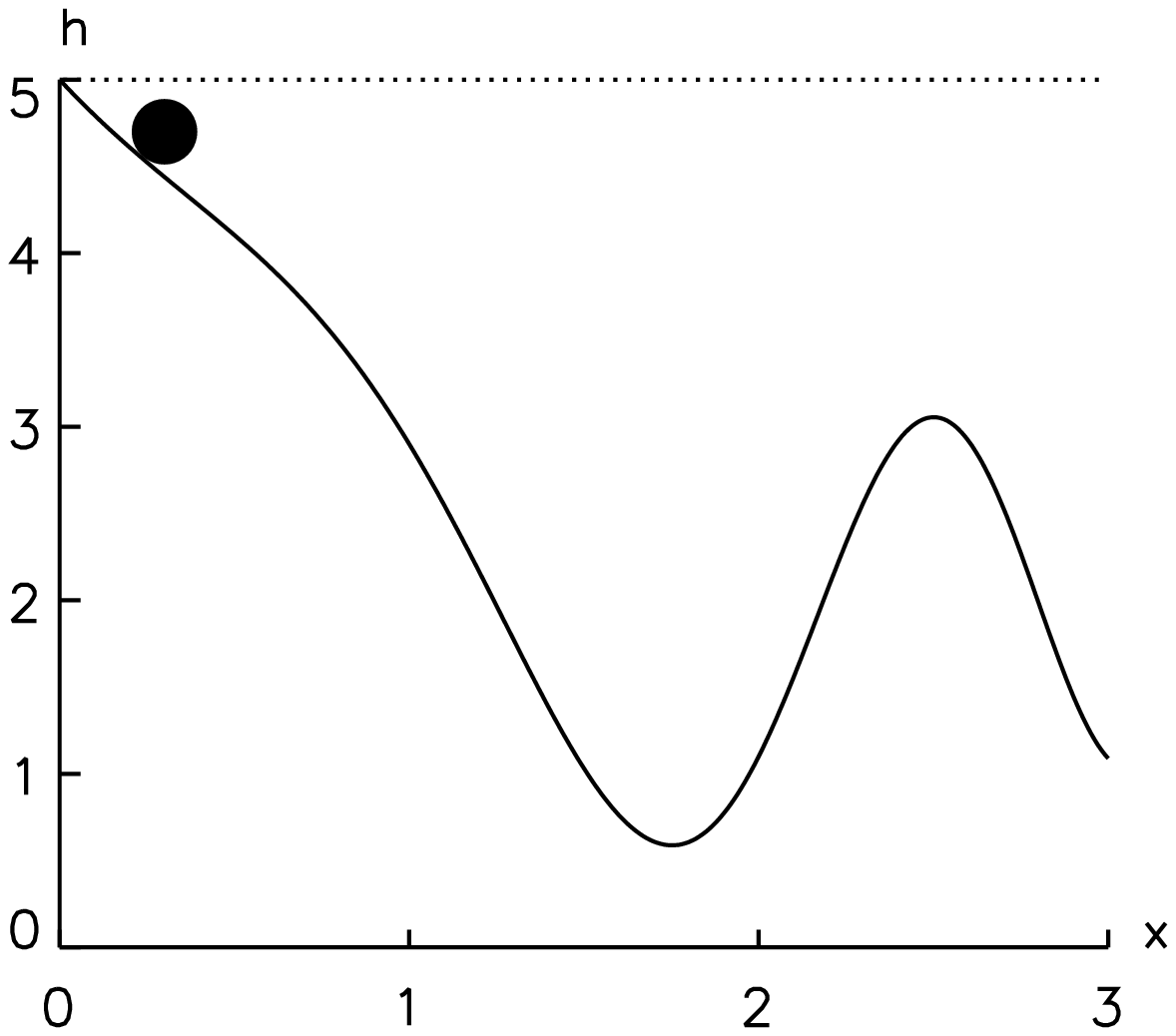}
\caption{\label{fig:primaryconfs_colorized}{\em Left panel:} Figure from \cite{Knop:2003iy}, differentiating
between regions expanding forever, recollapsing eventually, having no Big Bang, accelerating and decelerating, as a function of the densities in matter and a cosmological constant. Note that current constraints on $\om$ and $\ol$ as derived from Type Ia supernovae are significantly better \cite{Betoule:2014frx}.
{\rm Right panel:} A particle rolling frictionless along a track, maximally reaching the height indicated by the dotted line. The potential energy of the ball $U=mgh$ follows the track of the particle. When the potential energy decreases (i.e., the particle rolls down the track), the kinetic energy of the particle increases. When the potential energy increases (i.e., the particle rolls up the track), the kinetic energy decreases. In this paper, we show how the different regions of the left panel can be understood using a similar picture.}
\end{figure}

As an example, consider the left panel of figure~\ref{fig:primaryconfs_colorized}, frequently reproduced from \cite{Knop:2003iy} showing the observational evidence for an accelerating univeral expansion, first discovered in \cite{1998AJ....116.1009R,1999ApJ...517..565P} a few years earlier\footnote{A discovery subsequently awarded the Nobel prize in Physics in 2011.}. 

Although since long obsolete in terms of the actual confidence contours derived from observed distances to Type Ia supernovae, it also differentiates 
between regions labeled as expanding forever, recollapsing eventually, having no Big Bang, accelerating and decelerating.
For a student recently introduced to the Friedmann and acceleration equations, understanding how these regions come about can be challenging.

In the right panel of figure~\ref{fig:primaryconfs_colorized}, a small ball is pictured being released from rest at height $h=5$ at zero horizontal position, $x=0$. Assuming no friction is acting on the ball, it is easy to qualitatively get a full picture of the dynamics of the rolling ball. The total energy of the ball will be the constant sum of the kinetic and potential energy of the ball where the potential energy is given by $U=mgh(x)$. Here, $m$ is the mass of the ball, $g$ the gravitational acceleration at Earth and $h$ the (arbitrarily normalized) height of the ball. When the potential energy decreases, the kinetic energy increases and vice versa.

In the following, we will show how the different regions in the left panel can be trivially understood using the mechanical picture of a rolling ball depicted in the right panel. 

\section{Method}

Einstein's field equations with a homogeneous and isotropic metric ansatz give two differential equations for the time evolution of the scale factor $a(t)$ of the universe.
The first Friedmann equation is given by 
\be\label{eq:fe}
H_0^2\left(\frac{\dot{a}}{a}\right)^{2}\equiv H^{2}=\frac{8\pi G\rho}{3}+\frac{\Lambda}{3}-\frac{kc^2}{R_{0}^{2}}\frac{1}{a^{2}},
\ee
and the second, often denoted the acceleration equation, by 
\be\label{eq:acc}
H_0^2\frac{\ddot{a}}{a}=-\frac{4\pi G}{3}\left(\rho+\frac{3p}{c^2}\right)=-\frac{4\pi G\rho}{3}(1+3\omega).
\ee
Here, $\Lambda$ is a cosmological constant, $k=[-1,0,1]$, $R_{0}$ is the radius of curvature and $\rho$, $p$ and $\omega$ denote the total energy density, pressure and equation of state, respectively. The current value of the scale factor $a$ is normalized to unity. Dots denote derivatives with respect to the dimensionless time coordinate $\tau\equiv H_0 t$. $H_0$ can generally be any constant of dimension $t^{-1}$, but for models for which the Hubble parameter today is not zero, corresponds to this value, i.e. $H_0=H(a=1)$. From the Friedmann equations, we can derive the energy conservation equation  
\be\label{eq:ec}
\dot\rho+3\frac{H}{H_0}\left(\rho +\frac{p}{c^2}\right)=0,
\ee
or
\be\label{eq:rhoa}
\rho=\rho_0 a^{-3(1+\omega)},
\ee
where a subscript zero denotes the current value. Equations~(\ref{eq:ec}) and (\ref{eq:rhoa}) hold independently for all energy species as long as they do not convert into each other. So far, equations have been written in the standard textbook form. 

In order to make contact to the familiar picture of a particle rolling along a frictionless track, we define $\Omega\equiv8\pi G\rho/(3H_{0}^{2})$, $\ol\equiv\Lambda/(3H_0^2)$ and $\ok\equiv-kc^2/(H_{0}^{2}R_{0}^{2})$, and rewrite equation (\ref{eq:fe}) as 
\be\label{eq:adot2}
\dot{a}^{2}-\Omega a^2=\ok,
\ee
where, in a universe containing radiation, pressureless matter and a cosmological constant
\be
\Omega = \frac{\Omega_{R}}{a^{4}}+\frac{\om}{a^3}+\ol .
\ee
Here, $\Omega_{R}$ and $\om$ are the current, dimensionless energy densities in radiation and matter respectively.
Equation~\ref{eq:adot2} is the energy equation $K+U=E$ for a particle\footnote{The corresponding mass for the particle is $2$. Note that $K, U$ and $E$ are dimensionless.} moving one dimensionally along coordinate $a$ with kinetic energy $K\equiv{\dot a}^2$, potential energy $U\equiv -\Omega a^2$ and total energy $E\equiv \ok$. If $\dot a \neq 0$ today, that is corresponding to ${\dot a}=1$, the total energy is given by $E = \ok=1-\Omega_0$. For a static universe however, the total energy and curvature does not need to obey this relation. Taking the derivative of equation~\ref{eq:adot2} with respect to $\tau$, we obtain the acceleration equation in the form familiar from conservative systems in classical mechanics\footnote{The unfamiliar factor of $1/2$ is due to the fact that we did not define the kinetic energy as $\dot{a}^2/2$.}
\be
\ddot{a}=-\frac{1}{2}\frac{dU(a)}{da}.
\ee

In the standard model of the universe, containing radiation, pressureless matter and a cosmological constant, the potential and total energy and is given by 
\begin{eqnarray}   
U =-\left(\frac{\Omega_{R}}{a^{2}}+\frac{\om}{a}+\ol a^{2}\right),\\
E = \ok = 1-\Omega_R-\om-\ol.
\end{eqnarray}
All we have to do to understand the allowed expansion histories of a given model is to plot the potential
energy function. The expansion history will be given by the motion of a rolling particle that can maximally reach the height $\ok$. Since in general, energy densities are positive, the shape of the potential energy function has some general properties. For any energy density component with $w>-1/3$, the contribution to the potential energy will go to minus infinity as the scale factor goes to zero and to zero as the scale factor become infinitely large. For any energy density component with $w<-1/3$, the contribution to the potential energy will go to zero as the scale factor goes to zero and to minus infinity as the scale factor become infinitely large. 

If $w=-1/3$, the contribution to the potential energy will be constant. Since it will affect the total energy in the same way as the curvature term, any such energy component will not affect the expansion history of the universe, given the expansion velocity today\footnote{On the other hand, as noted in the section~\ref{sec:intro}, it will affect cosmological observations through the geometrical curvature that enters into distance measures.}. 
The only exception to energy densities being positive is the cosmological constant ($\omega =-1$) that could have any sign. The contribution to the potential from a positive cosmological constant will go to minus infinity as the scale factor become infinitely large whereas the potential contribution from a negative cosmological constant will go to infinity in the same limit.

\section{Dynamical models}\label{sec:dyn}

\begin{figure}
\includegraphics[width=8cm]{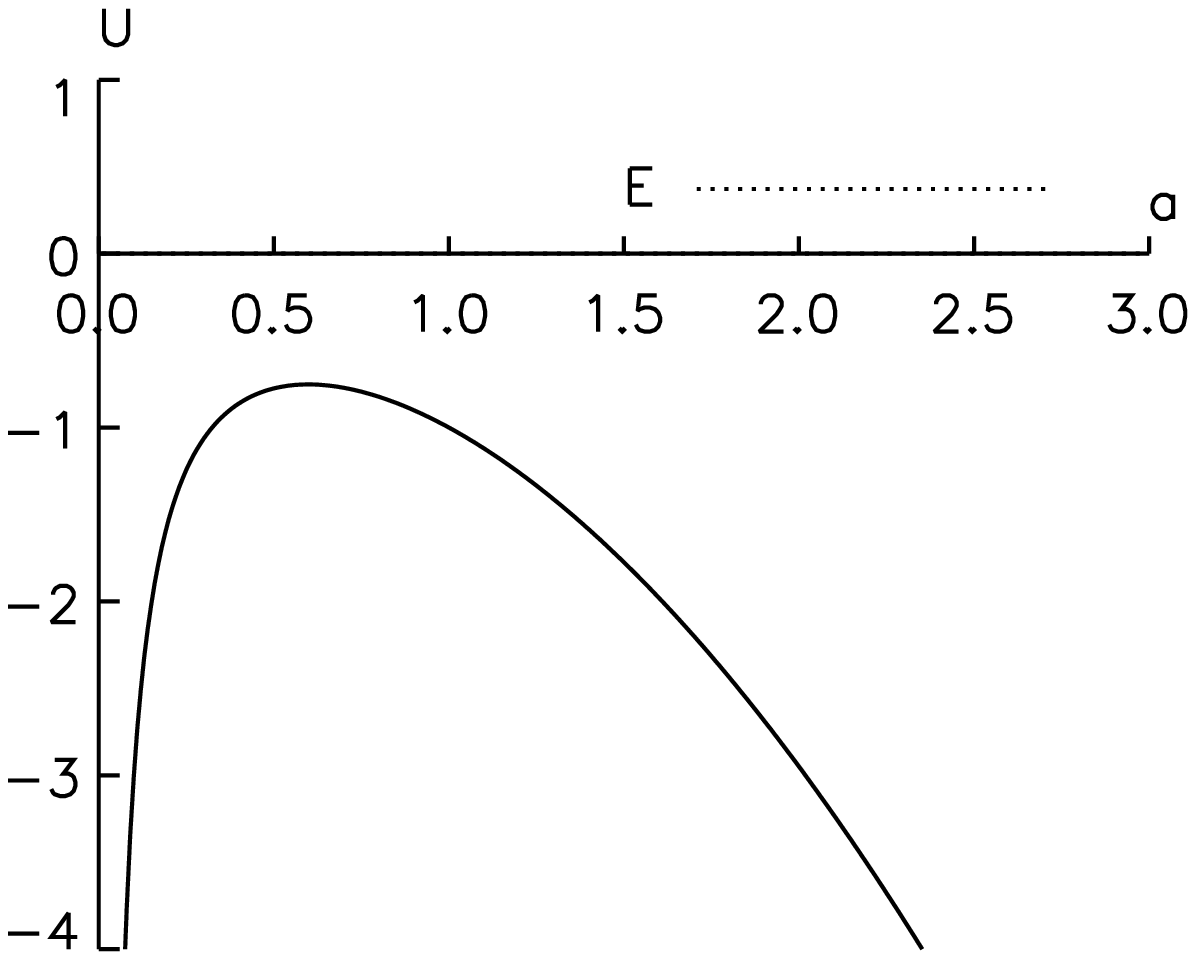}\includegraphics[width=8cm]{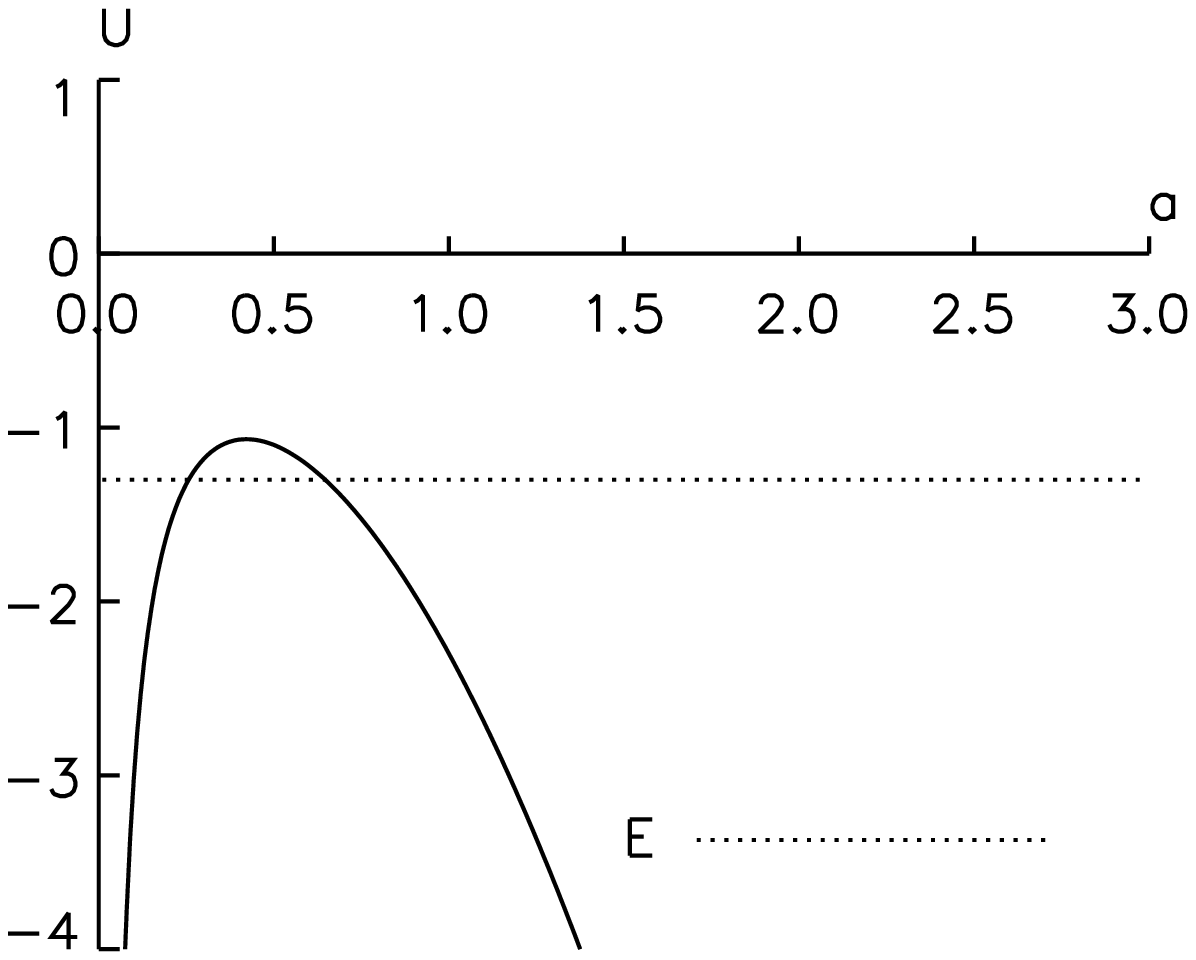}
\caption{\label{fig:M=03L=07}{\em Left panel:} The concordance cosmological model with $\om= 0.3$, $\ol= 0.7$ and zero spatial curvature. Our expansion history is represented by a particle rolling in from the left. Since $a_0=1$, we are currently living in a period of accelerated expansion. {\em Right panel:} The potential energy function (solid line) of a model with $\om= 0.3$ and $\ol=2$ and total energy (dotted line) of $E=1-\om-\ol=-1.3$. In this case we have two different solutions: First, the expansion history is constrained to $a\gesim 2/3$ and the model does not have a Big Bang, i.e., $a=0$. Second, a universe originating from a Big Bang will expand up to $a\sim 1/3$ after which it will contract again.}
\end{figure}

We will first study the case of the so called concordance model, see e.g. \cite{Ade:2015xua}. It has $\om= 0.3$, $\ol= 0.7$ and zero spatial curvature, and thus zero total energy. The potential energy function is shown in the left panel of figure~\ref{fig:M=03L=07}. Note that since the total energy is zero, it will be hidden by the abscissa. The expansion history can now easily be understood as a particle rolling in from the left up the slope with decreasing velocity, corresponding to the decelerating matter dominated period. It rolls over the hill at $a\sim 0.5$ where the velocity is at its (non-zero) minimum and starts rolling down the slope with ever increasing velocity. This corresponds to the current accelerated phase when the cosmological constant is dominating the energy content of the Universe. Since the scale factor then grows indefinitely while the matter density approaches zero, this is sometimes described as the Universe approaching a Big Chill.  

If we increase the value of the cosmological constant to, e.g. $\ol=2$, we get the case depicted in the right panel of figure~\ref{fig:M=03L=07}. The total energy, indicated by the dotted line, is $E=1-\om-\ol=-1.3$. At $a=1$, we are either in a state of decelerated contraction if the particle is rolling up the slope from to the left, or in a state of accelerated expansion if the particle is rolling down the slope to the right after turning around at $a\sim 2/3$. In either case, the particle will never reach $a=0$ and the corresponding universe does not have a Big Bang.
The region to the left with $0\lesim a\lesim 1/3$ corresponds to a universe originating from a Big Bang, expanding up to $a\sim 1/3$ and then starting to contract again. Compared to the case of $\om=0.3$ and $\ol=0.7$ we thus have the somewhat counter intuitive result that increasing the value of the cosmological constant can counteract the universal expansion and even make it reverse. This is due to the fact that the total energy of the system is lowered.  

\begin{figure}
\includegraphics[width=8cm]{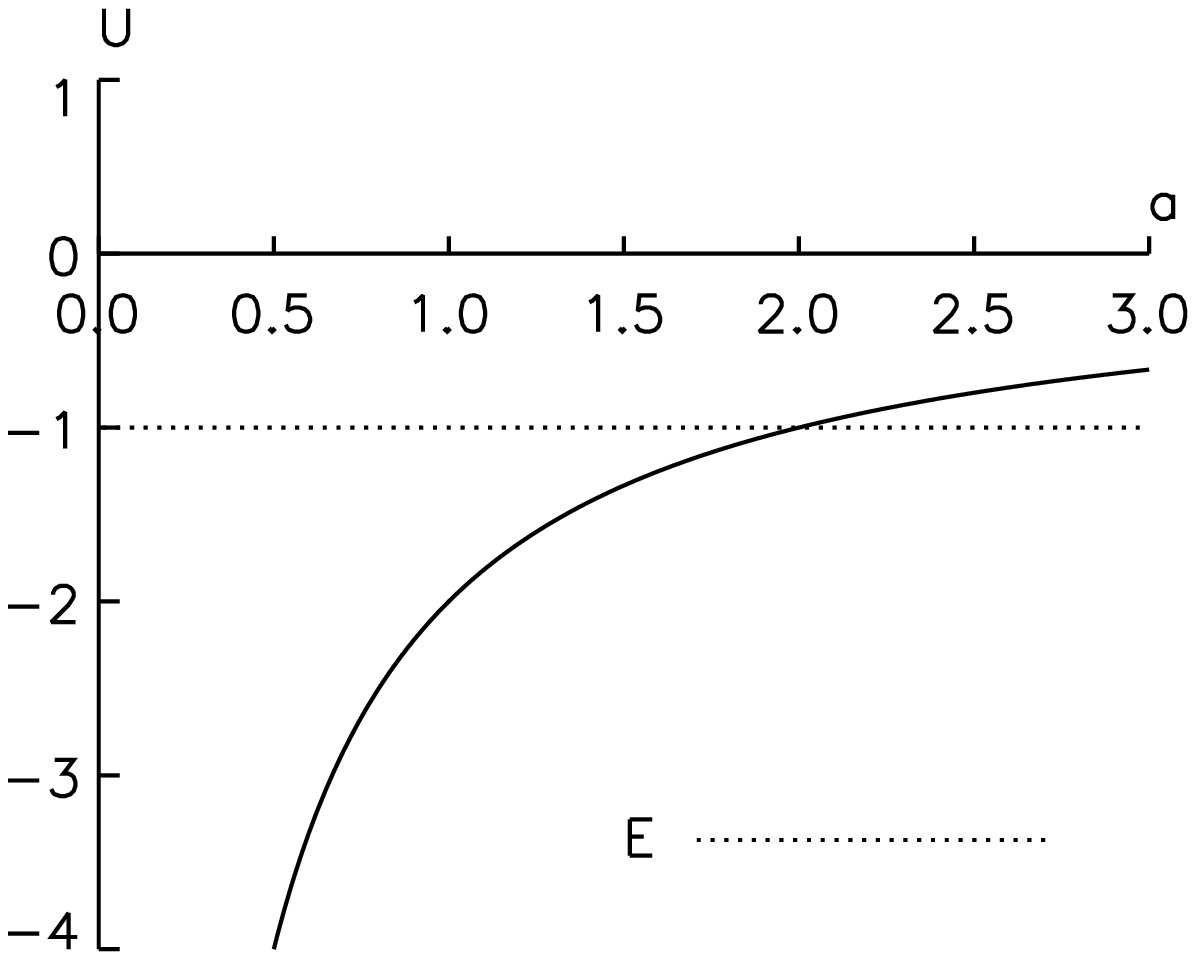}\includegraphics[width=8cm]{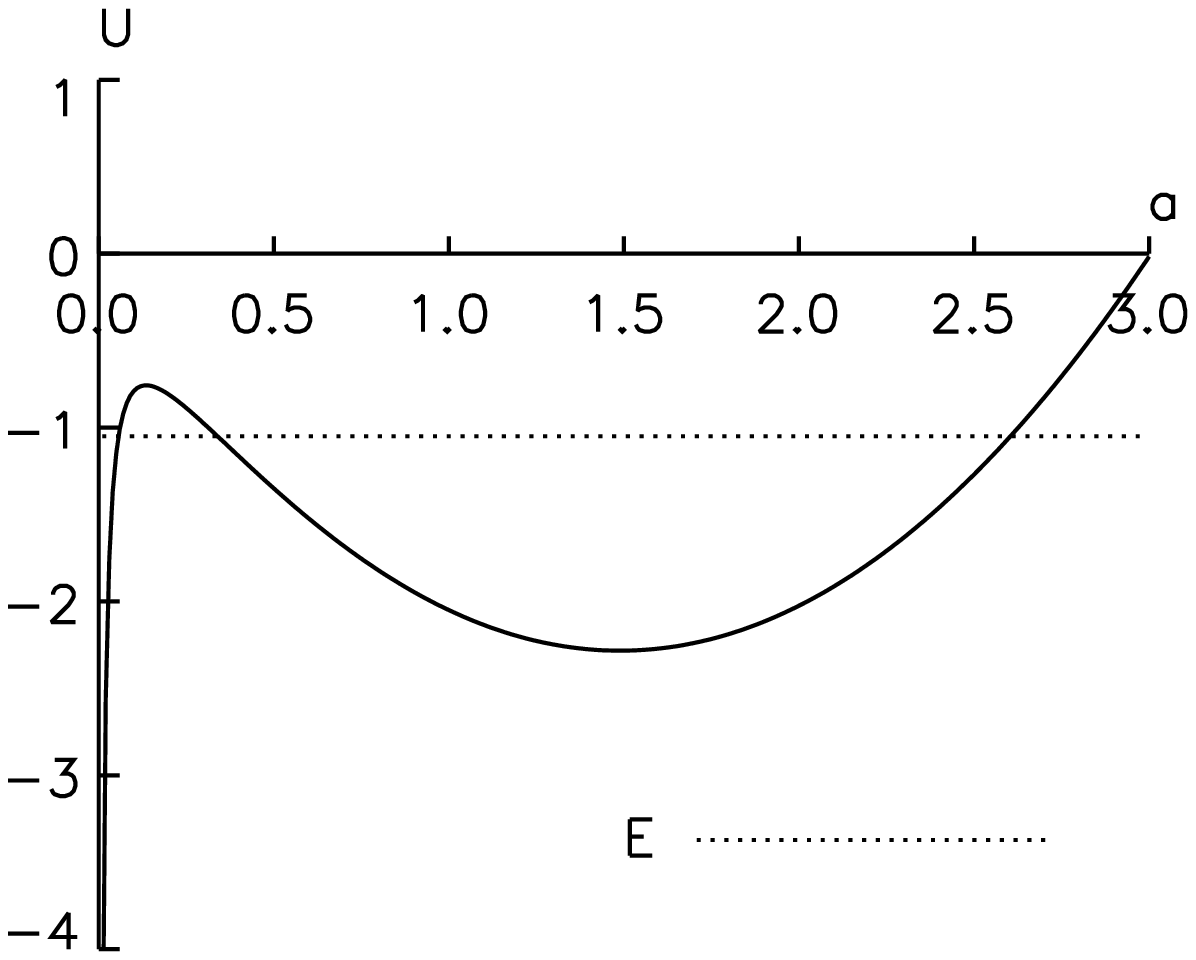}
\caption{\label{fig:M=2}{\em Left panel:} The potential energy (solid line) for a matter dominated overclosed universe with $\om= 2$ and total energy $E_t=1-\om=-1$ (dotted line). The expansion history is represented by a particle rolling in from the left. At $a=2$, the kinetic energy becomes zero and the particle starts rolling back towards a Big Crunch. {\em Right panel:} The potential energy (solid line) for a matter density of $\om=0.05$,  a negative cosmological constant $\ol=-1$, and $\Omega_X=3$ with $\omega_X=-2/3$. The universe will oscillate back and forth between $a\sim 0.3$ and $a\sim 2.5$}
\end{figure}

In the left panel of figure~\ref{fig:M=2}, we depict the case of a matter dominated overclosed universe with $\om=2$ and negative total energy $E=1-\om=-1$. A particle rolling in from the left will reach $a=2$ at which the velocity is zero and the particle starts rolling back again. This corresponds to the case of a universe first expanding with ever decreasing velocity and then contracting with ever increasing velocity down to a Big Crunch. 

Including a negative cosmological constant will necessarily make the universe enter a contracting phase at some point\footnote{The exception is if there is a positive energy component with $\omega<-1$ or $\omega=-1$ and larger absolute value of the energy density compared to the cosmological constant. In principle, if the potential has a saddle point where the total energy is zero, entering a contracting phase can also be avoided.}. An interesting possibility is shown in the right panel of figure~\ref{fig:M=2}, where in addition to a matter density of $\om=0.05$ and a negative cosmological constant $\ol=-1$, we have added a component $\Omega_X=3$ with $\omega_X=-2/3$, corresponding to a large energy fraction in, e.g. domain walls. In this case, we will have a universe that oscillates back and forth between a minimum scale factor $a\sim 0.3$ and a maximum value $a\sim 2.5$. 

\section{Static solutions}\label{sec:stat}
\begin{figure}
\includegraphics[width=8cm]{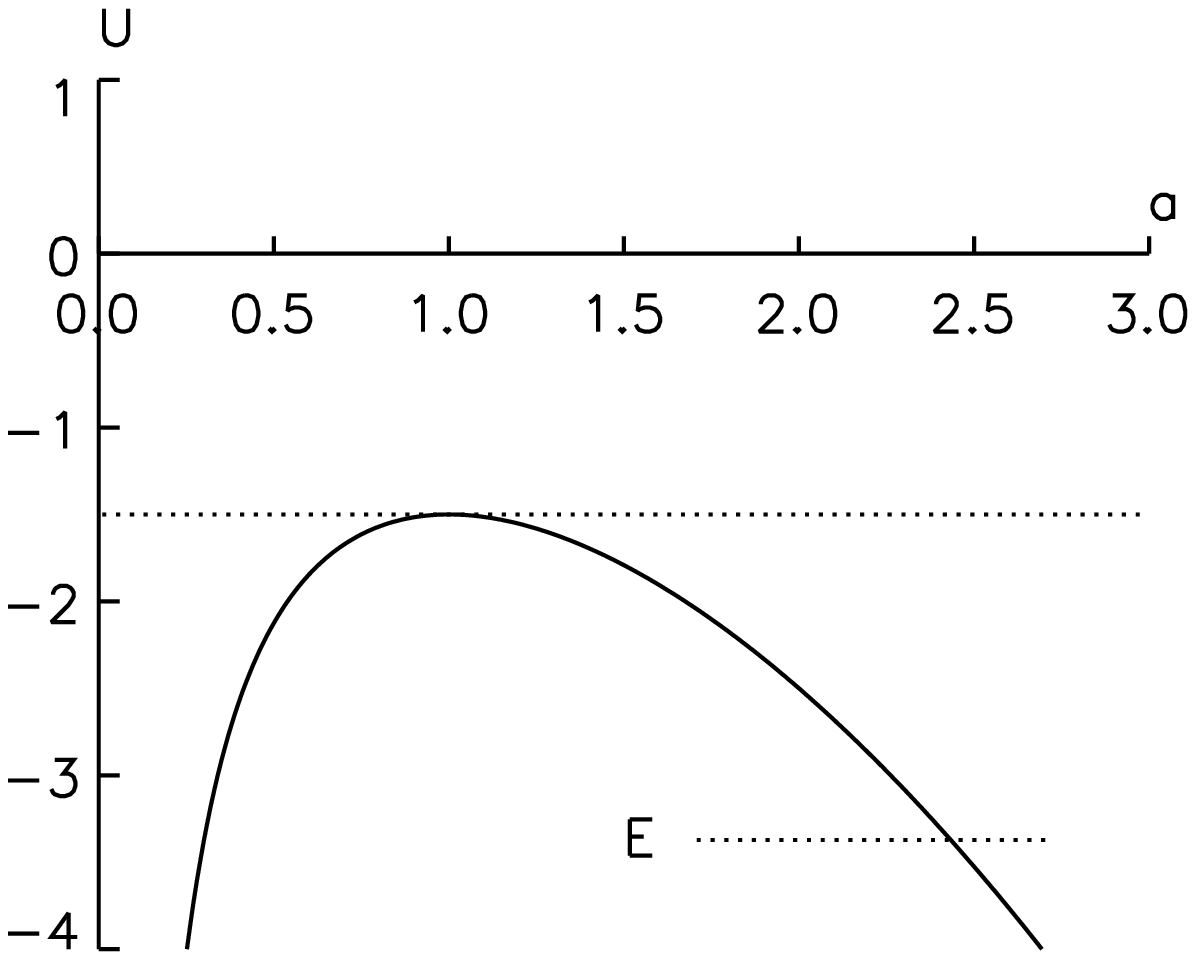}\includegraphics[width=8cm]{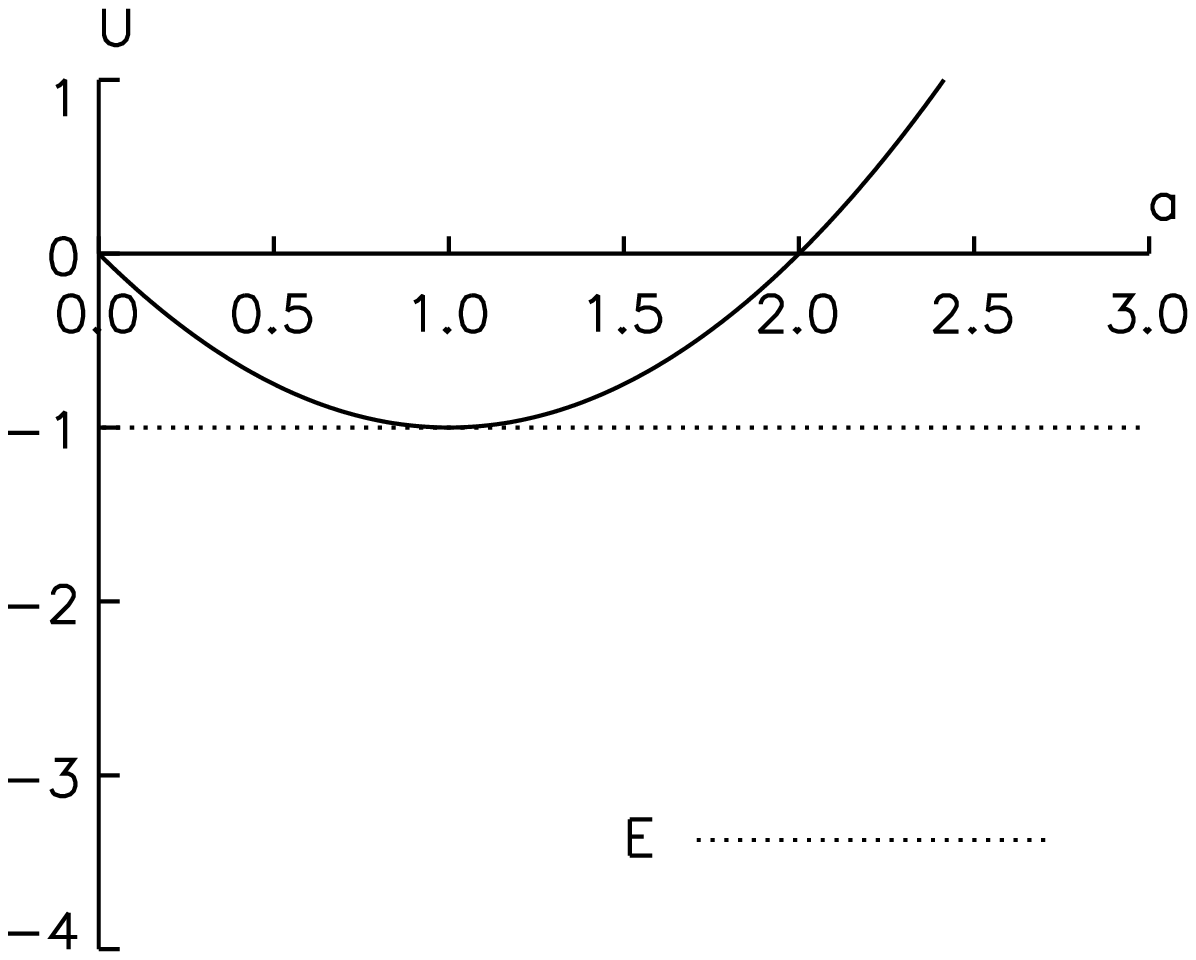}
\caption{\label{fig:es}{\em Left panel:} The potential (solid line) and total (dotted line) energy for Einsteins (unstable) static universe with $\ol = -\ok /3 = \om /2$.  {\em Right panel:} The potential (solid line) and total (dotted line) energy for a stable static universe dominated by domain walls with energy density $\Omega_X=2$ and equation of state $\omega_X=-2/3$ and a negative cosmological constant given by $\ol = \ok = -\Omega_X /2$.}
\end{figure}

Einstein first introduced the cosmological constant in order to find static solutions for the scale factor $a$ with $\dot a=\ddot a=0$. In a universe with matter, $\om$, and a cosmological constant, $\ol$, this corresponds to 
\be
\ol = -\frac{\ok}{3} = \frac{\om}{2},
\ee
as shown in the left panel of figure~\ref{fig:es}. It is obvious that this is not a stable situation; the slightest perturbation and the particle will start rolling down the potential, either to left or the right depending on the nature of the perturbation.  

The only way to obtain a stable static solution is to make the static point a minimum in the potential. This can be accomplished, e.g. as in figure~\ref{fig:es}, by having zero matter density, a component $\Omega_X=2$ with $\omega_X=-2/3$ corresponding to a dominating domain wall structure, and a negative cosmological constant given by  
\be
\ol = \ok = -\frac{\Omega_X}{2}.
\ee
If perturbed, the solution will oscillate around the minimum of the potential.

\section{Summary and conclusion}
In this paper, it is argued that a simple mechanical picture of the background expansion allows us to obtain a full, qualitative understanding of possible expansion histories given the energy density components of the universe.   
The cases depicted in sections~\ref{sec:dyn} and \ref{sec:stat} nicely illustrates how this can be done trivially also for models with quite complicated energy contents, yielding such diverse behaviour as oscillations between expansion and contraction as well as stable static solutions.

How then can the regions in the left panel of figure~\ref{fig:primaryconfs_colorized}  from \cite{Knop:2003iy} be trivially understood, as advertised in the introduction? 

The division between open and closed models is determined by the total energy, $E$, being positive (open) or negative (closed). Except for the concordance model in the left panel of figure~\ref{fig:M=03L=07}, all models discussed in this paper have closed spatial geometries.

Whether the universe is accelerating or decelerating is determined by if the particle is rolling upwards a slope (decelerating) or downwards (accelerating). Note that the dashed line in figure~\ref{fig:primaryconfs_colorized} denotes the division line determined by the state of the universe today. In the mechanical picture, we can trivially determine the dynamical state at any given redshift. Also note that any model displaying acceleration at some redshift also would have deceleration at the same point would the direction of the particle be reversed, that is if the direction of the expansion is reversed and vice versa. 

Whether the universe will expand forever (e.g. the concordance model) or recollapse eventually (e.g. a matter dominated universe as in the left panel of figure~\ref{fig:M=2}) will be determined by if the potential energy is larger than the total energy at some redshift.

Finally, the universe will not have a Big Bang if the rolling particle is confined to a region not including $a=0$, e.g. as in the solution with $a\gesim 2/3$ in the right panel of figure~\ref{fig:M=03L=07}.

Although the dynamics of our Universe probably follow quite closely the motion of a particle rolling from the left to the right of the left panel in figure~\ref{fig:M=03L=07}, the mechanical picture outlined here allows for a full understanding of the evolution of universes with arbitrary energy content. Hopefully, this picture can give both a simpler and a deeper understanding of the Friedmann equation and its possible solutions. 

\ack I acknowledge support for this study from the Swedish Research Council. Also, I am grateful to Jonas Enander and the referees for careful readings of the manuscript and useful suggestions. 
  
\appendix
\section{Inflation}
\begin{figure}
\centering
\includegraphics[width=8cm]{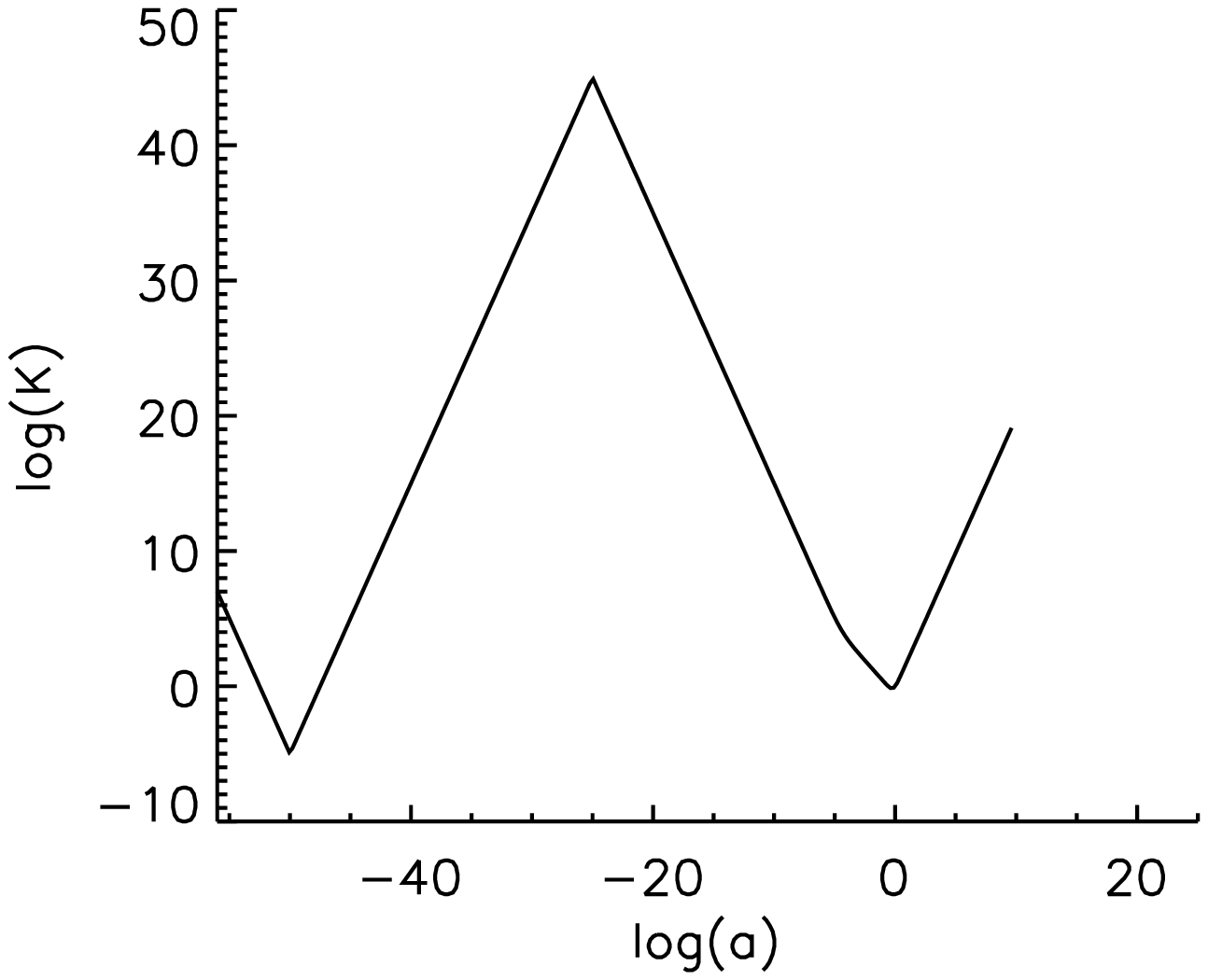}
\caption{\label{fig:Inflation} The kinetic energy of the universal expansion as a function of the scale factor. We assume that the Universe was dominated by radiation at $a<10^{-50}$ and that inflation took place at $10^{-50}<a<10^{-25}$. We live at $\log a = 0$. Note that the kinetic energy today can be comparable to that at the onset of inflation.}
\end{figure}

In the inflationary scenario \cite{1978AnPhy.115...78B,1979JETPL..30..682S,1981PhRvD..23..347G,1982PhLB..108..389L}, at some point in the early Universe, the energy density was dominated by a component, $\Omega_i$ behaving similarly to the cosmological constant, say between scale factors $a_b$ and $a_e$. During this epoch, the potential energy function was given by 
\begin{eqnarray}   
U \sim -\Omega_i a^{2},\\
E \sim 0,
\end{eqnarray}
after which some mechanism converted the energy density in $\Omega_i$ to radiation and matter. We need approximately 60 e-foldings of inflation in order to successfully obtain the initial conditions for the subsequent universal evolution, for which the theory was devised. The scale factor will then increase by a factor of $\sim 10^{25}$ during inflation, the potential energy will decrease by a factor of $\sim 10^{50}$ and the kinetic energy will increase by the same factor. Assuming that inflation took place at $10^{-50}<a<10^{-25}$ and that the Universe before that was dominated by a radiation like component\footnote{This assumption of course being completely hypothetical.}, the kinetic energy of the expansion as a function of the scale factor is depicted in figure~\ref{fig:Inflation}. Since the observable Universe (given by the particle horizon), today has a radius of $\sim 4\cdot 10^{26}$ m, at the end of inflation, the corresponding radius was $\sim 40$ m, and at the start of inflation $\sim 4\cdot 10^{-24}$ m, or $\sim 3\cdot 10^{11}\, l_P$. In this picture, at $a\sim 10^{-56}$, we reach the Planck density and we therefore do not extend the plot to smaller values of the scale factor.

\bibliography{bibliography}{}

\providecommand{\newblock}{}
\begin{thebibliography}{10}
\expandafter\ifx\csname url\endcsname\relax
  \def\url#1{{\tt #1}}\fi
\expandafter\ifx\csname urlprefix\endcsname\relax\def\urlprefix{URL }\fi
\providecommand{\eprint}[2][]{\url{#2}}

\bibitem{Einstein:1917ce}
Einstein A 1917 {\em Sitzungsber. Preuss. Akad. Wiss. Berlin (Math. Phys.)\/}
  {\bf 1917} 142--152

\bibitem{deSitter:1917zz}
de~Sitter W 1917 {\em Mon. Not. Roy. Astron. Soc.\/} {\bf 78} 3--28

\bibitem{Einstein:1915ca}
Einstein A 1915 {\em Sitzungsber. Preuss. Akad. Wiss. Berlin (Math. Phys.)\/}
  {\bf 1915} 844--847

\bibitem{Realdi}
Realdi M and Peruzzi G 2009 {\em General Relativity and Gravitation\/} {\bf 41}
  225--247 ISSN 0001-7701
  \urlprefix\url{http://dx.doi.org/10.1007/s10714-008-0664-y}

\bibitem{Lemaitre:1927zz}
Lemaitre G 1927 {\em Annales Soc. Sci. Brux. Ser. I Sci. Math. Astron. Phys.\/}
  {\bf A47} 49--59

\bibitem{Friedmann:1924bb}
Friedmann A 1924 {\em Z. Phys.\/} {\bf 21} 326--332 [Gen. Rel.
  Grav.31,2001(1999)]

\bibitem{Way:2013ky}
Way M~J 2013 {\em ASP Conf. Ser.\/} {\bf 471} 97 (\textit{Preprint}
  \eprint{1301.7294})

\bibitem{Szydlowski:2006ma}
Szydlowski M and Hrycyna O 2007 {\em Annals Phys.\/} {\bf 322} 2745--2775
  (\textit{Preprint} \eprint{astro-ph/0602118})

\bibitem{Padmanabhan:2006kz}
Padmanabhan T 2006 {\em AIP Conf. Proc.\/} {\bf 843} 111--166
  (\textit{Preprint} \eprint{astro-ph/0602117})

\bibitem{Sonego:2011rb}
Sonego S and Talamini V 2012 {\em Am. J. Phys.\/} {\bf 80} 670--679
  (\textit{Preprint} \eprint{1112.4319})

\bibitem{Knop:2003iy}
Knop R~A {\em et~al.\/} (Supernova Cosmology Project) 2003 {\em Astrophys.
  J.\/} {\bf 598} 102 (\textit{Preprint} \eprint{astro-ph/0309368})

\bibitem{1998AJ....116.1009R}
{Riess} A~G, {Filippenko} A~V, {Challis} P, {Clocchiatti} A, {Diercks} A,
  {Garnavich} P~M, {Gilliland} R~L, {Hogan} C~J, {Jha} S, {Kirshner} R~P,
  {Leibundgut} B, {Phillips} M~M, {Reiss} D, {Schmidt} B~P, {Schommer} R~A,
  {Smith} R~C, {Spyromilio} J, {Stubbs} C, {Suntzeff} N~B and {Tonry} J 1998
  {\em Astronomical Journal\/} {\bf 116} 1009--1038 (\textit{Preprint}
  \eprint{astro-ph/9805201})

\bibitem{1999ApJ...517..565P}
{Perlmutter} S, {Aldering} G, {Goldhaber} G, {Knop} R~A, {Nugent} P, {Castro}
  P~G, {Deustua} S, {Fabbro} S, {Goobar} A, {Groom} D~E, {Hook} I~M, {Kim} A~G,
  {Kim} M~Y, {Lee} J~C, {Nunes} N~J, {Pain} R, {Pennypacker} C~R, {Quimby} R,
  {Lidman} C, {Ellis} R~S, {Irwin} M, {McMahon} R~G, {Ruiz-Lapuente} P,
  {Walton} N, {Schaefer} B, {Boyle} B~J, {Filippenko} A~V, {Matheson} T,
  {Fruchter} A~S, {Panagia} N, {Newberg} H~J~M, {Couch} W~J and {Project} T~S~C
  1999 {\em Astrophysical Journal\/} {\bf 517} 565--586 (\textit{Preprint}
  \eprint{astro-ph/9812133})

\bibitem{Betoule:2014frx}
Betoule M {\em et~al.\/} (SDSS) 2014 {\em Astron. Astrophys.\/} {\bf 568} A22
  (\textit{Preprint} \eprint{1401.4064})

\bibitem{Ade:2015xua}
Ade P~A~R {\em et~al.\/} (Planck) 2015  (\textit{Preprint} \eprint{1502.01589})

\bibitem{1978AnPhy.115...78B}
{Brout} R, {Englert} F and {Gunzig} E 1978 {\em Annals of Physics\/} {\bf 115}
  78--106

\bibitem{1979JETPL..30..682S}
{Starobinski{\v i}} A~A 1979 {\em Soviet Journal of Experimental and
  Theoretical Physics Letters\/} {\bf 30} 682

\bibitem{1981PhRvD..23..347G}
{Guth} A~H 1981 {\em Physical Review D\/} {\bf 23} 347--356

\bibitem{1982PhLB..108..389L}
{Linde} A~D 1982 {\em Physics Letters B\/} {\bf 108} 389--393

\end{thebibliography}

\end{document}